\documentclass[sigconf]{acmart}
\setcopyright{none}
\renewcommand\footnotetextcopyrightpermission[1]{}
\AtBeginDocument{%
  }

\usepackage{booktabs} 
\usepackage{siunitx}  
\usepackage{makecell}
\usepackage{graphicx}
\usepackage{subcaption}
\usepackage{float}
\usepackage{comment}
\usepackage{xcolor}
\usepackage{placeins}

\definecolor{autocolor}{RGB}{0, 100, 150}     

\begin{document}

\title{WebKnoGraph: GNN-Powered Internal Linking} 

\author{Emilija Gjorgjevska}
\email{emilia.gjorgjevska@tum.de}
\affiliation{%
  \institution{Technical University of Munich}
  \city{Heilbronn}
  \country{Germany}
}

\author{Georgina Mirceva}
\email{georgina.mirceva@finki.ukim.mk}
\affiliation{%
  \institution{Ss. Cyril and Methodius University in Skopje}
  \city{Skopje}
  \country{North Macedonia}
}

\author{Miroslav Mirchev}
\email{miroslav.mirchev@finki.ukim.mk}
\affiliation{  \institution{Ss. Cyril and Methodius University in Skopje}  \city{Skopje}  \country{North Macedonia}}

\renewcommand{\shortauthors}{Gjorgjevska et al.}

\begin{abstract}
Internal link optimization is a recurring task in search engine optimization, yet many production workflows rely on manual judgment, fixed page templates, or generic tool recommendations. Practitioners need ways to evaluate candidate links before deployment because link changes can redistribute authority and affect semantic coherence in ways that are difficult to isolate after release. We present \textsc{WebKnoGraph}, an open-source framework for evaluating internal linking strategies on website crawls. The framework models a website as a directed graph, represents pages by embeddings, scores candidate links with GraphSAGE, and evaluates interventions by embedding the site into larger host environments. We instantiate \textsc{WebKnoGraph} on a production crawl of Kalicube.com and compare automatic with expert-assisted link selection in an empirical FineWeb-based host graph and a synthetic Barabási--Albert host graph, using PageRank-based authority metrics and semantic coherence. The results show that automatic selection generally produces stronger authority redistribution, with higher Authority Yield, but also larger semantic coherence costs. Expert-assisted selection better preserves semantic coherence and, when targeting low-PageRank pages, achieves the highest Authority Yield, although with the least favorable loss--gain balance. Authority Volatility provides an additional stability perspective, but is interpreted cautiously because the two regimes use different numbers of intervention sets. These findings support a practical workflow in which candidate intervention sets are generated at scale, evaluated jointly across authority gain, volatility, loss--gain balance, and semantic coherence, and then reviewed for editorial deployability before implementation.
\end{abstract}



\thanks{M.M. is also affiliated with the Complexity Science Hub, Vienna, Austria. E.Gj. was also affiliated with WordLift and Kalicube.}


\maketitle

\section{Introduction}
\label{sec:introduction}

Search visibility has direct economic and reputational value, making search engine optimization (SEO) a high-impact applied domain \cite{ledford2015search,mordor_seo_market_2025,lewandowski2021influence,malaga2008worst}. 
Internal links shape how search engines discover pages, interpret site structure, and distribute authority across a website \cite{takumiseo}. 
Yet internal linking remains difficult to evaluate systematically in production workflows.

The core challenge is operational. 
Teams add links through templates, navigation elements, editorial modules, and manual cross-linking, but must decide which pages should receive authority, which pages should act as donors, and whether proposed links are semantically appropriate and feasible to implement. 
As websites grow, these decisions become harder because site structures often evolve without strategic planning, content can become difficult to discover \cite{gjorgjevska2021content}, and manual optimization becomes costly \cite{lvivuniseo}.

PageRank and related link analysis methods offer a formal model of authority propagation \cite{brin1998anatomy,page1999pagerank}, but this theory is rarely translated into reproducible internal-link evaluation workflows. 
In practice, teams often rely on manual judgment, fixed rules, or generic recommendations, with limited ability to compare candidate interventions before deployment. 
Live evaluation is also noisy: deployed link changes interact with content updates, external links, crawl behavior, seasonality, ranking-system changes, and signals such as relevance, anchor context, freshness, topical consistency, and spam safeguards \cite{fetterly2004spam,yandex_ranking_factors,googlesearchfactors2025,smaili2024future}. 
This makes it difficult to attribute observed outcomes to a specific internal-link intervention.

We introduce \textsc{WebKnoGraph} \cite{webknograph}, an open-source framework for evaluating internal-linking strategies on production website crawls before implementation. 
The framework models a website as a directed graph, uses semantic content embeddings and GraphSAGE-based link identification to propose candidate links, and evaluates interventions by embedding the site into larger host graphs. 
This provides a controlled graph-based analysis for comparing candidate strategies using measurable indicators of authority gain, volatility, loss-gain balance, and semantic coherence.

We instantiate \textsc{WebKnoGraph} on a production crawl of Kalicube.com and compare automatic with expert-assisted link selection in two host environments: an empirical FineWeb-based graph and a synthetic Barabási--Albert graph. 
Rather than claiming direct ranking or traffic effects, we provide reproducible pre-deployment evidence on the structural and semantic tradeoffs of candidate internal-link interventions.

In this paper, we make four contributions: 
(i) we formulate internal link optimization as an applied graph intervention problem grounded in production SEO workflows; 
(ii) we release \textsc{WebKnoGraph}, a reproducible open-source framework for evaluating candidate internal links before deployment; 
(iii) we evaluate automatic and expert-assisted strategies on a production website crawl embedded into empirical and synthetic host graphs; and 
(iv) we derive practitioner-facing lessons about tradeoffs between authority redistribution, volatility, loss-gain balance, and semantic coherence.

\section{Related work}
\label{sec:related-work}

Work on single-site internal-link rewiring under realistic SEO and editorial constraints remains limited. 
Our approach connects two research strands: PageRank optimization through link-level interventions, and learning-based link prediction with graph learning.

PageRank and related authority models provide the theoretical basis for link-based ranking \cite{brin1998anatomy,page1999pagerank}. 
Prior work has studied how link structure can be manipulated or optimized under different constraints. 
Baeza-Yates et al. \cite{baeza2005pagerank} analyze collusive link structures, showing that low-ranked pages can benefit from coordinated linking, while high-ranked pages face diminishing returns. 
Avrachenkov and Litvak \cite{avrachenkov2004decomposition} decompose PageRank into core and dangling contributions, providing strategies for redistributing authority flow. 
Csáji, Jungers, and Blondel \cite{csaji2014pagerank} formulate PageRank maximization as an edge-selection problem, while Ohsaka et al. \cite{ohsaka2018boosting} study heuristics for boosting PageRank with a limited number of internal links. 
These works motivate link intervention as an optimization problem, but they do not directly address production websites where links must also satisfy semantic, editorial, and implementation constraints.

Search-engine practice and patent literature further indicate that link evaluation extends beyond raw graph topology. 
Prior work and filings describe link-context weighting, freshness-sensitive scoring, graph proximity measures, template abstraction, and spam-resistant link evaluation \cite{Anderson_2025,US8577893B1,US8407231B2,US9165040B1,US20080134015A1}. 
Sundin \cite{Sundin2025} argues that AI-driven search shifts retrieval toward answer-oriented systems, increasing the need to model interactions between content structure, semantic interpretation, and authority distribution. 
This supports the need for internal-link evaluation methods that account for both structural authority flow and semantic coherence.

Learning-based link prediction provides the second foundation of our work. 
Hyperlink prediction has been explored for Wikipedia links, news-to-Wikipedia suggestions, and blog networks \cite{gerlach2021multilingual,lyu2018real,wu2011future}. 
Earlier approaches relied on topological heuristics or shallow graph embeddings, while graph neural networks provide more expressive models that combine graph structure with node attributes. 
Representative architectures include Graph Convolutional Networks \cite{kipf2016semi}, Graph Attention Networks \cite{velivckovic2017graph}, and GraphSAGE \cite{hamilton2017inductive}, within the broader graph neural networks literature \cite{khoshraftar2024survey}. 

\textsc{WebKnoGraph} builds on these foundations by applying GNN-based link identification to real website graphs and evaluating proposed internal links through PageRank-based graph analysis in both empirical and synthetic host environments. 
This bridges link-prediction methods with production-crawl SEO evaluation, where candidate links must be assessed not only for authority redistribution, but also for semantic coherence and editorial feasibility.


\section{Problem Statement}
\label{sec:problem-statement}

We consider the problem of optimizing the internal link structure of a website in order to improve the distribution of authority across its pages. The target website is modeled as a directed site graph \(G_S~=~(V_S,~E_S)\), where nodes \(V_S\) represent pages and edges \(E_S~\subseteq~V_S~\times~V_S\) represent existing internal hyperlinks.

\textit{Intervention model.}
An internal-link boosting intervention consists of adding a set of directed links \(E_k \subseteq (V_S \times V_S) \setminus E_S\) to the site graph, yielding an augmented site graph \(G_S' = (V_S, E_S \cup E_k)\). The intervention size is fixed at approximately 240 links to reflect realistic editorial constraints; we denote this intervention budget by \(k\). Intervention links are selected from a candidate link space in either the automatic or expert-assisted regime.

\textit{Objective.}
Let \(\Delta \mathrm{PR}_i\) denote the relative change in PageRank of site page \(i \in V_S\) induced by an intervention. The objective of internal-link optimization is to identify interventions that improve authority across site pages under this fixed intervention budget.

\textit{Hosting and external context.}
Because page authority on the Web is externally influenced, the effect of internal-link interventions cannot be assessed in isolation. We therefore model the website within a broader context by embedding the site graph \(G_S\) into an external host graph \(G_H=(V_H,E_H)\) and introducing a set of directed bridging edges \(E_B~\subseteq~(V_S~\times~V_H)~\cup~(V_H~\times~V_S)\), representing links between the site and the surrounding web. This yields a baseline composite graph \(G_C=(V_C,E_C)=(V_S~\cup~V_H,~E_S~\cup~E_H~\cup~E_B)\) and an intervention composite graph \(G_{C'}=(V_C,~E_C~\cup~E_k)\) including the added internal links. Authority is computed on these composite graphs, while all reported outcomes are restricted to site pages \(V_S\).

\textit{Evaluation.}
Intervention effects are evaluated by measuring PageRank changes of individual pages and aggregating them into overall metrics. We assess (i) the average authority gain normalized by the intervention budget
(\emph{Authority Yield}), (ii) the stability of this gain across repeated embeddings into the host environment (\emph{Authority Volatility}), (iii) the ratio between authority losses and gains across site pages (\emph{Authority Down/Up Ratio}), and (iv) the change in semantic coherence induced by the added links (\emph{Semantic-Coherence change}). Formal definitions are provided in Section~\ref{sec:methodology-eval-framework}.

\section{Methodology}
\label{sec:methodology}

\begin{figure}[p] 
    \centering
    \includegraphics[height=0.95\textheight, keepaspectratio,trim=2.5cm 0cm 2.5cm 0cm,clip]{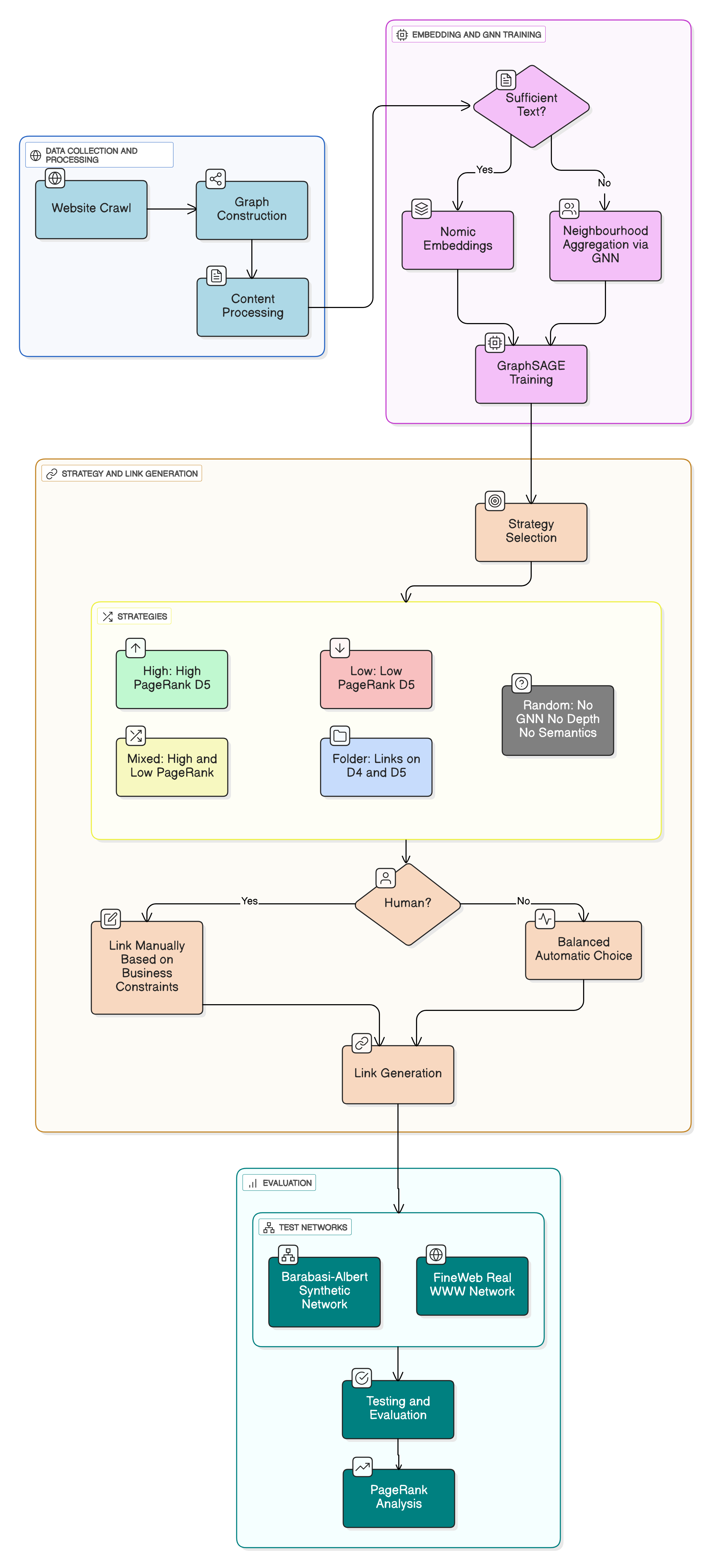}
    \caption{A flowchart of the WebKnoGraph system design}
    \label{fig:flowchart}
\end{figure}

This section outlines the multi-faceted methodology designed to investigate the effect of internal linking on website architecture and authority. Figure~\ref{fig:flowchart} summarizes the four-stage pipeline that guides the organization of this section. We present the data and graph representation, the modeling and algorithm used to generate link recommendations (including feature construction via content embeddings), the internal linking strategies, the human-in-the-loop review protocol, and the evaluation framework definitions.

\subsection{Data}
\label{sec:methodology-data}

\textit{Kalicube data.}
Our primary dataset is a full production crawl of \texttt{Kalicube.com}, conducted in Q2 2025. The crawl comprises approximately $2{,}000$ English pages, of which $1{,}841$ are incorporated into the site graph after URL normalization and parsing. The website 
architecture has clearly defined topical sections and a stable editorial structure. For each page, we extract the internal hyperlink neighborhood and the full textual content, and we compute directory depth from the URL path, assigning the homepage depth~0. This representation supports analysis of authority redistribution induced by controlled internal-link interventions.

\textit{FineWeb dataset.}
To place the Kalicube site graph into a broader web-scale context, we draw on the FineWeb dataset~\cite{penedo2024the}, a large-scale web corpus designed for research use and derived from multiple Common Crawl snapshots. Using streaming access to the \texttt{HuggingFaceFW/fineweb} dataset, we collect $200{,}000$ training records from the \texttt{sample\_10BT} configuration. For each record, we retain the fields \texttt{url} and \texttt{text}. Records with missing content, malformed URLs, or non-HTTP(S) schemes are removed. The resulting collection of FineWeb pages provides an empirical basis for constructing a realistic external web environment, as described below.

\subsection{Host environments}
\label{sec:methodology-host}

To evaluate internal-link interventions under different network conditions, we use two host environments: an empirical network derived from FineWeb and a synthetic Barabási--Albert (BA) network. These environments provide structurally distinct contexts in which the same site graph and intervention strategies are evaluated.

\textit{FineWeb-based host environment.}
From the 200,000 sampled FineWeb pages described in Section~\ref{sec:methodology-data}, we construct a directed page-level host graph with 500,000 edges. The graph combines four edge-generation mechanisms: (i) explicit URL references extracted from page text ($\sim$40\%), which capture direct references to other pages; (ii) links between pages with similar URL-path structure ($\sim$20\%), which approximate section-level navigation and topical neighborhoods; (iii) intra-domain links ($\sim$30\%), which approximate menus, breadcrumbs, and internal cross-references; and (iv) random links ($\sim$10\%), which introduce controlled topological noise. All edges are normalized, merged, and deduplicated. The final FineWeb-based host contains 224,242 unique page identifiers and 500,000 directed edges. While node identifiers correspond to real pages, the edge set should be interpreted as a controlled empirical proxy for web connectivity rather than as a reconstruction of the Web.

\textit{Barabási--Albert synthetic host environment.}
As a complementary baseline, we generate a synthetic host graph using the Barabási--Albert preferential-attachment model~\cite{barabasi1999emergence}. The BA model produces scale-free networks with heterogeneous degree distributions and hub nodes, properties commonly associated with the World Wide Web. The graph contains $N=100{,}000$ nodes and $m=2$ links per new node, resulting in approximately 200,000 directed edges. This environment abstracts away content- and domain-specific effects while providing a controlled reference topology.

\subsection{Graph-based candidate-link scoring}
\label{sec:methodology-model}
This stage turns page content and graph structure into candidate donor--target links.

\textit{Content embeddings.} 
During data preparation, boilerplate elements such as navigation menus, footers, and advertisements are removed with trafilatura~\cite{barbaresi2021trafilatura}. The remaining text is embedded using nomic-embed-text-v1~\cite{nussbaum2024nomic}, with task prefixes for page and anchor-text representations, producing 768-dimensional, L2-normalized vectors. We select this model because it provides open weights, training code, and data, while achieving retrieval performance comparable to widely used proprietary embedding models. These content embeddings serve as input features for the graph encoder.

\textit{Graph-contextual node embeddings.}
We use the GraphSAGE architecture~\cite{hamilton2017inductive} because it supports scalable and inductive representation learning on large graphs through neighborhood sampling and aggregation. In our implementation, a two-layer mean-aggregation model combines page-level content features with information from immediate and second-order graph neighborhoods. Sampling a fixed-size set of neighbors at each layer keeps training feasible on large host graphs. Operating in an inductive setting, GraphSAGE can generate embeddings for new or modified pages from features and local neighborhoods, without retraining the full model. The resulting node embeddings encode topical information and structural context, including for pages with sparse textual content.

\textit{Generating candidate links.} 
The model scores potential donor--target links by combining global graph topology, local neighborhood structure, and content similarity, so that the resulting rankings reflect structural link opportunities rather than semantic similarity alone. These rankings are used within the strategy-specific candidate link spaces defined in Section~\ref{sec:methodology-strategies}.

\subsection{Internal linking strategies}
\label{sec:methodology-strategies}
We define five strategies that construct candidate link spaces targeting different aspects of website structure. The \emph{High} strategy focuses on boosting high-PageRank pages at directory depth~5, leveraging established authority distribution principles. The \emph{Low} strategy boosts low-PageRank pages at the same depth to enhance underperforming content visibility. The \emph{Mixed} strategy combines both high- and low-performing pages at depth~5 for simultaneous optimization across performance tiers. The \emph{Folder} strategy targets pages within depths~4 and~5 to improve connectivity among mid-level content. Finally, the \emph{Random} strategy serves as an unguided baseline, defining a candidate link space without restrictions based on PageRank, semantic relevance, directory depth, or other strategic criteria.

\subsection{Intervention-set construction}
\label{sec:methodology-intervention-construction}
For each strategy and selection regime, an intervention set follows the intervention budget \(k\) defined above. In all experiments, this intervention size is chosen to reflect a realistic editorial budget, while allowing minor deviations in the realized updated graphs when selected links are duplicates or already present in the baseline site graph. The automatic and expert-assisted regimes operate within the candidate link space defined by each strategy, allowing us to examine how the mode of selecting links affects the resulting intervention.

\textit{Automatic selection.}
In the automatic regime, GraphSAGE ranks candidate donor--target pairs within the strategy-specific candidate link space. Candidate links are filtered to remove self-links and existing links, and are constrained by the relevant strategy criteria, such as PageRank tier, directory depth, or folder membership. To avoid concentrating added links on a small number of pages, load-balancing lowers the priority of pages that have already contributed or received several selected links. Links are then selected iteratively until the intervention set reaches the target budget.

\textit{Expert-assisted selection.}
In the expert-assisted regime, WordLift SEO professionals select links from the GraphSAGE-ranked candidate link space defined by each strategy. Two intermediate-level experts conduct the primary selection, with oversight and validation by a senior expert. Their selection criteria include page-template compatibility, directory or section alignment, semantic relevance, user-experience implications, and operational implementation constraints. The resulting intervention sets therefore incorporate deployment-related considerations that are not explicitly optimized in the automatic selection procedure.

For the \emph{Random} strategy, boosted pages and added links are selected without GraphSAGE ranking or expert relevance assessment: the selection is performed algorithmically at random in the automatic regime and manually at random in the expert-assisted regime. Different intervention sets within the same strategy and regime capture variation in the selected donor--target links. This design enables comparison between automatic and expert-assisted selection while holding the intervention budget and the strategy-defined candidate link spaces constant.

\subsection{Evaluation framework}
\label{sec:methodology-eval-framework}

We evaluate intervention effects by measuring how authority is redistributed within the site graph when it is evaluated in a broader network context. For each intervention, we compare PageRank outcomes on the baseline and intervention composite graphs in both the FineWeb-based and BA host environments defined in Section~\ref{sec:methodology-host}. This allows identical internal-link interventions to be evaluated under structurally distinct external network conditions, rather than on the site graph alone. Intervention effects are summarized using four configuration-level metrics that capture complementary dimensions of performance: \emph{Authority Yield}, \emph{Authority Volatility}, \emph{Authority Down/Up Ratio}, and \emph{Semantic-Coherence Change}. All reported outcomes concern the target site graph \(G_S\).

\textit{Relative PageRank change.}
For each page \(i \in V_S\) and simulation run \(r\), we compute the relative PageRank change between the baseline \(G_{C,r}\) and intervention \(G_{C',r}\) composite graphs as
\[
\Delta \mathrm{PR}_{i,r}(\%) =
\frac{\mathrm{PR}_i(G_{C',r}) - \mathrm{PR}_i(G_{C,r})}
{\mathrm{PR}_i(G_{C,r})}
\times 100 .
\]
We then summarize page-level changes within each run by \(\mu_r\), defined as the mean of \(\Delta \mathrm{PR}_{i,r}\) over all site pages, and summarize the overall effect by \(\bar{\mu}\), defined as the average of \(\mu_r\) across all runs.

\textit{Authority Yield.}
Authority Yield (AY) measures the marginal effectiveness of an intervention, normalized by the intervention budget \(k\), and is defined as
\begin{equation}
\mathrm{AY} =
\frac{\bar{\mu}}{k}.
\end{equation}
This normalization isolates the efficiency of a linking strategy relative to the intervention budget, enabling meaningful comparison across configurations.

\textit{Authority Volatility.}
Authority Volatility (AV) captures the stability of Authority Yield across simulation runs:
\begin{equation}
\mathrm{AV} =
\frac{\sigma_{\mu}}{k},
\end{equation}
where \(\sigma_{\mu}\) denotes the standard deviation of \(\mu_r\) across runs. Low volatility indicates more predictable authority gains, whereas high volatility signals sensitivity to stochastic variation in graph embedding or link placement.

\textit{Authority Down/Up Ratio}
To characterize the balance between authority losses and gains, we analyze the direction of page-level changes. Using a neutrality threshold \(\tau = 0.025\%\), we define \(U_r\) and \(D_r\) as the number of site pages in run \(r\) with \(\Delta\mathrm{PR}_{i,r} > \tau\) and \(\Delta\mathrm{PR}_{i,r} < -\tau\), respectively. Let \(\bar{U}\) and \(\bar{D}\) denote the corresponding averages across runs. Authority Down/Up Ratio can be defined as
\begin{equation}
\mathrm{R_{D/U}} =
\frac{\bar{D}}{\max(\bar{U},1)},
\end{equation}
where values below one indicate that authority gains dominate losses, while values larger than one indicate the opposite.

\textit{Semantic-Coherence.} Let \(\mathbf{x}_i\) be the L2-normalized content embedding of page \(i\). For a site graph \(G_S = (V_S, E_S)\), semantic coherence is the average cosine similarity across internal links:
\[
\mathrm{SC}(G_S) =
\frac{1}{|E_S|}
\sum_{(i,j)\in E_S}
\mathbf{x}_i^\top \mathbf{x}_j .
\]
Higher values indicate links between more topically related pages, while lower values indicate weaker content-level support. The reported intervention effect is the run-averaged change
\[
\Delta \mathrm{SC} =
\frac{1}{R}\sum_{r=1}^{R}
\left[
\mathrm{SC}(G_{S',r}) - \mathrm{SC}(G_S)
\right],
\]
where \(G_{S',r}\) denotes the realized updated site graph for run \(r\), containing approximately \(k \approx 240\) added internal links. Positive values indicate improved coherence, while negative values indicate reduced coherence.

\section{Results}
\subsection{Experimental Setup}

We compare five internal-linking strategies---High, Low, Mixed, Folder, and Random---under automatic and expert-assisted selection regimes. For each strategy and selection regime, intervention sets are constructed under the same intervention budget \(k\), corresponding to approximately 240 added internal links. We evaluate these interventions in the two host environments defined in Section~\ref{sec:methodology-host}: a synthetic Barabási--Albert (BA) graph and an empirical FineWeb-based graph. The complete Kalicube.com site graph with 1{,}841 pages serves as the fixed intervention target.

Each intervention set is evaluated through repeated bridging simulations. Our primary analysis uses the middle bridging range, \([35,65]\), representing moderate external connectivity. As an additional analysis, we evaluate the lower and higher ranges, \([5,35]\) and \([65,95]\), to examine how outcomes vary with external connectivity. In each simulation, the selected intervention set remains fixed while the external environment varies. We keep the total simulation budget constant across selection regimes. For automatic selection, each strategy has 10 intervention sets, each evaluated with 100 bridging simulations, yielding \(10 \times 100 = 1{,}000\) simulations per strategy and range. For expert-assisted selection, each strategy has 2 intervention sets, corresponding to the two independently selected expert link batches, and each set is evaluated with 500 bridging simulations, yielding \(2 \times 500 = 1{,}000\) simulations per strategy and range. 
Because the two selection regimes use different numbers of intervention sets, AV is primarily used as a stability indicator within each regime rather than to compare volatility between regimes.

\begin{figure*}[t]
\centering
\includegraphics[width=\textwidth]{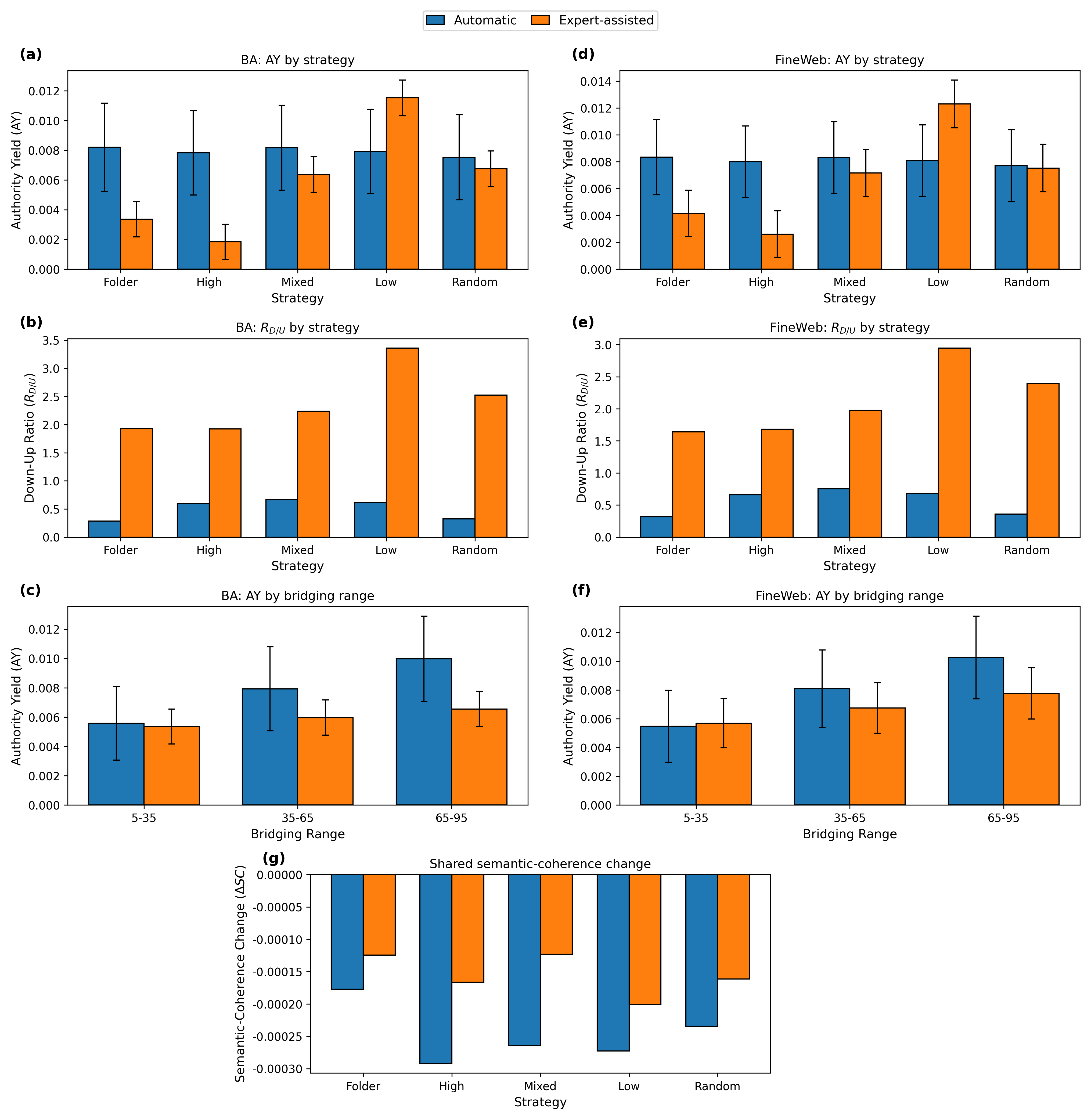}
\caption{Authority-flow and semantic-coherence effects by host environment, strategy, and selection regime. Panels (a)--(c) report results for the BA host environment, while panels (d)--(f) report the corresponding FineWeb results. Panels (a), (b), (d), and (e) use the representative 35--65 bridging range. Panels (c) and (f) report Authority Yield across bridging ranges, averaged across strategies, with Authority Volatility shown as error bars. Panel (g) reports Semantic-Coherence Change ($\Delta SC$), which is computed on the augmented site graph and is therefore shared across host environments.}
\label{fig:results_summary_1}
\end{figure*}

\subsection{Strategy-Level Effects}
\label{sec:strategy-effects}

We first examine the strategy-level effects of the interventions in Figure~\ref{fig:results_summary_1}. Panels~(a) and~(b) report the Barabási--Albert results, while panels~(d) and~(e) report the corresponding FineWeb results. These panels report the primary analysis based on the middle bridging range, $[35,65]$. Semantic-Coherence Change is reported in panel~(g).

Because Semantic-Coherence Change is computed on the augmented site graph rather than on the composite host graph, the $\Delta SC$ values are independent of the BA and FineWeb host environments. The host environment affects the PageRank-based metrics, but not the semantic similarity between connected site pages.

Across both host environments, the automatic strategies show relatively small differences in AY compared with the larger separation observed for $\mathrm{R_{D/U}}$. This suggests that, within the automatic regime, the strategies differ less in their average authority gain relative to the intervention budget than in how authority gains and losses are distributed across the site. Similar AY values can therefore correspond to different loss--gain profiles: one strategy may achieve a comparable average gain while affecting fewer pages negatively, whereas another may produce a similar gain with a less favorable balance between pages gaining and losing authority.

The expert-assisted regime shows a different AY pattern. In both host environments, the expert-assisted \textit{Low} strategy achieves the highest strategy-level AY. However, it also has the least favorable $\mathrm{R_{D/U}}$ profile, indicating that its stronger average authority gain is accompanied by a less favorable distribution of gains and losses across site pages. Under expert-assisted selection, \textit{Folder} and \textit{High} produce more favorable $\mathrm{R_{D/U}}$ values, whereas \textit{Mixed} and \textit{Random} fall between these extremes.

Under automatic selection, \textit{Folder} produces the most favorable $\mathrm{R_{D/U}}$ profile in both host environments, while \textit{Random} also remains comparatively favorable. The \textit{High} strategy does not achieve a better loss--gain balance than \textit{Random} when the final donor--target links are selected automatically. Under expert-assisted selection, however, \textit{High} achieves a more favorable $\mathrm{R_{D/U}}$ profile than \textit{Random} in both host environments. This suggests that the strategic objective of boosting high-authority pages is sensitive to the final link choices made within the candidate space. Expert-assisted selection appears to translate the \textit{High} strategy into a more favorable loss--gain profile, indicating that expert review can be important when translating a strategy-defined candidate space into the final intervention set.

The semantic-coherence results in panel~(g) add a further dimension to this comparison. All intervention configurations produce small negative changes in semantic coherence, indicating that the added links slightly reduce the average semantic similarity among connected page pairs. This is expected when the internal link structure is expanded beyond the original site architecture. Nevertheless, the magnitude of the decrease differs across configurations. Automatic interventions generally introduce larger negative $\Delta SC$ values than their expert-assisted counterparts, suggesting that expert review helps preserve semantic coherence. Among the automatic strategies, \textit{Folder} produces the smallest semantic penalty, whereas \textit{High}, \textit{Mixed}, and \textit{Low} introduce larger losses.

Taken together, the strategy-level results show that no single metric is sufficient for evaluating an internal-link intervention. The expert-assisted \textit{Low} strategy achieves the highest strategy-level AY, but this gain is accompanied by the least favorable $\mathrm{R_{D/U}}$ profile. Conversely, configurations with more moderate authority gains may better preserve the distribution of authority across pages or the semantic coherence of the link structure. Since these qualitative patterns are largely preserved across both host environments, the principal difference between the Barabási--Albert and FineWeb settings appears to be the magnitude of the observed effects rather than the underlying strategy-level tradeoffs.

\subsection{Automatic vs. Expert-Assisted Selection}
\label{sec:selection-regimes}

We next examine the aggregate difference between automatic and expert-assisted selection across bridging ranges. Panels~(c) and~(f) of Figure~\ref{fig:results_summary_1} report AY averaged across strategies for each bridging range, with Authority Volatility (AV) shown as error bars. Unlike the strategy-level comparison above, this analysis summarizes the average authority-redistribution effect of each selection regime as external connectivity varies.

Across both host environments, automatic interventions generally produce higher average AY than expert-assisted interventions, and this difference becomes more pronounced at higher bridging ranges. This indicates that, when averaged across strategies, automatic selection has stronger authority-redistribution potential as the site is embedded into increasingly connected host environments. However, this aggregate pattern does not imply that automatic selection is preferable for every strategy or evaluation objective. As shown in the strategy-level analysis, the expert-assisted \textit{Low} configuration achieves the highest AY at the primary bridging range, although with the least favorable $\mathrm{R_{D/U}}$ profile.

The AV error bars are generally smaller for expert-assisted than for automatic selection in both host environments. However, because the two selection regimes use different numbers of intervention sets, this pattern is reported descriptively and is not interpreted as evidence that expert-assisted selection is inherently more stable.

Overall, the results indicate that automatic selection is effective for generating interventions with stronger average authority gains, whereas expert assistance better preserves semantic coherence and can produce competitive or superior AY in specific strategy-level settings. The results therefore support a workflow in which automatic generation is combined with multi-objective evaluation and expert review, rather than treating either selection regime as uniformly preferable.

\section{Limitations}
\label{sec:limitations}

The framework relies on link structure and page content extracted from a permitted production crawl and from web-corpus data.
This makes the approach reproducible and usable without privileged access to analytics systems, but excludes behavioral and operational signals such as clicks, impressions, conversions, crawl frequency, and server logs.

We do not evaluate the system through live A/B testing on production websites. 
Because SEO environments change continuously through content updates, external links, technical changes, and algorithm updates, our results should be interpreted as production-crawl evaluation analyses of authority redistribution and semantic coherence, not as direct measurements of ranking or traffic impact.

Finally, the empirical evaluation is limited to a single production domain. 
Although interventions are tested in both empirical and synthetic host graphs, further validation across domains such as e-commerce, news, and documentation sites is needed to assess broader robustness.

\section{Conclusion and discussion}
\label{sec:conclusion}

We presented \textsc{WebKnoGraph}, a graph-based framework for evaluating internal-link interventions before deployment. 
The framework combines content embeddings, GraphSAGE-based link identification, controlled intervention strategies, and PageRank-based evaluation in empirical and synthetic host environments.

Our results show that internal linking should be treated as a multi-objective graph intervention problem. 
\textsc{WebKnoGraph} exposes tradeoffs between authority redistribution, volatility, loss-gain balance, and semantic coherence, helping practitioners to compare candidate links before live SEO outcomes become difficult to attribute. 
These findings support a practical workflow in which candidate intervention sets are generated at scale, evaluated jointly across complementary criteria, and then reviewed for editorial deployability before implementation.

\section{Future work}
\label{sec:future-work}

We will extend \textsc{WebKnoGraph} in three directions. 
First, we will validate the framework on other production websites, including e-commerce, news, and documentation sites, to test whether the observed tradeoffs generalize across architectures and editorial constraints. 
Second, we will connect \textsc{WebKnoGraph} to operational SEO workflows, including the planned integration with WordLift’s Internal Links solution \cite{wordliftCreatingInternal}. 
This will support recommendation generation, expert review, rollout, monitoring, and practitioner feedback.
Third, we will extend production-crawl evaluation toward monitored live rollout and broaden interventions from link addition to link removal and pruning. 
Where attribution conditions allow, monitoring may include crawl frequency, indexed pages, impressions, clicks, engagement, conversions, and other business-relevant indicators.

Together, these extensions will move \textsc{WebKnoGraph} from pre-deployment evaluation toward an operational decision-support system for internal-link management, in which recommendations can be generated, evaluated, deployed, and monitored.

\section*{Data and Code Availability}
The WebKnoGraph framework, experiment scripts, and configuration files are available at
\url{https://github.com/martech-engineer/WebKnoGraph}.
We also release the Kalicube.com crawled dataset. The evaluation relies on the FineWeb dataset \cite{penedo2024the}, available at
\url{https://huggingface.co/datasets/HuggingFaceFW/fineweb}.
Synthetic Barabási–Albert graphs were generated with standard procedures and are reproducible with the released code.

\begin{acks}
M.M. was partially funded by the Austrian Ministry for Innovation, Mobility and Infrastructure through CSH's PostDoc Program. The authors thank WordLift and Kalicube for sharing data and contextual information and support relevant to this study.
\end{acks}

\bibliographystyle{ACM-Reference-Format}
\bibliography{bibliography}

@article{googlesearchfactors2025,
  title = {Leaked Google Search Algorithm Ranking Factors Database: By GrowthSRC Media},
  author = {{GrowthSRC Media}},
  journal = {searchrankingfactors.com},
  year = {2025},
  note = {Accessed: 12 September 2025},
  url = {https://searchrankingfactors.com/}
}

@misc{Anderson_2025,
  author    = {Shaun Anderson},
  title     = {Strategic SEO},
  publisher = {Hobo Web},
  year      = {2025},
  url       = {https://www.hobo-web.co.uk/strategic-seo-2025/}
}

@article{Sundin2025,
  title={Theorising notions of searching, (re)sources and evaluation in the light of generative AI},
  author={Sundin, Olof},
  journal={Information Research},
  volume={30},
  number={CoLIS},
  pages={291--302},
  year={2025},
  month={May},
  doi={10.47989/ir30CoLIS52258},
  url={https://publicera.kb.se/ir/article/view/52258}
}

@article{takumiseo, 
  title     = "Optimizing website ranking using long-tail keywords and internal
               linking: A case study",
  author    = "Sanjaya, Memet and Putra, Rizaldi and Utama, Deni and Prayoga,
               Arif",
  journal   = "jidt",
  publisher = "PT. Seulanga System Publisher",
  pages     = "31--36",
  month     =  aug,
  year      =  2025
}

@article{lvivuniseo,
  title     = "Evaluation of the effectiveness of offline search optimization
               in the {SEO} toolbox",
  author    = "{Ivan Franko Lviv National University} and Marchuk, O I and
               Kushnir, T M",
  journal   = "Mark. Digit. Technol.",
  publisher = "Odessa Polytechnic National University",
  volume    =  8,
  number    =  4,
  pages     = "44--57",
  month     =  dec,
  year      =  2024
}

@misc{webknograph,
author = {Gjorgjevska, Emilija and Miroslav Mirchev and Georgina Mircheva},
title = {WebKnoGraph: AI-Driven Framework for Large-Scale Internal Link Optimization},
url = {https://github.com/martech-engineer/WebKnoGraph},
year = {2024}
}

@inproceedings{
  penedo2024the,
  title={The FineWeb Datasets: Decanting the Web for the Finest Text Data at Scale},
  author={Guilherme Penedo and Hynek Kydl{\'\i}{\v{c}}ek and Loubna Ben allal and Anton Lozhkov and Margaret Mitchell and Colin Raffel and Leandro Von Werra and Thomas Wolf},
  booktitle={The Thirty-eight Conference on Neural Information Processing Systems Datasets and Benchmarks Track},
  year={2024},
  url={https://openreview.net/forum?id=n6SCkn2QaG}
}

@article{smaili2024future,
  title={The Future of Search Attention: Leveraging AI to Enhance PageRank's Influence},
  author={Smaili, Hasnae Amnoun$^1$ Naoual and Barboucha$^1$, Hamza and Kodad, Mohcine},
  journal={Advances in Smart Medical, IoT \& Artificial Intelligence: Proceedings of ICSMAI'2024, Volume 1},
  volume={11},
  pages={125},
  year={2024},
  publisher={Springer Nature}
}

@misc{nussbaum2024nomic,
      title={Nomic Embed: Training a Reproducible Long Context Text Embedder}, 
      author={Zach Nussbaum and John X. Morris and Brandon Duderstadt and Andriy Mulyar},
      year={2024},
      eprint={2402.01613},
      archivePrefix={arXiv},
      primaryClass={cs.CL}
}

@inproceedings{barbaresi2021trafilatura,
  title={Trafilatura: A web scraping library and command-line tool for text discovery and extraction},
  author={Barbaresi, Adrien},
  booktitle={Proceedings of the 59th Annual Meeting of the Association for Computational Linguistics and the 11th International Joint Conference on Natural Language Processing: System Demonstrations},
  pages={122--131},
  year={2021}
}

@inproceedings{gjorgjevska2021content,
  title={Content Engineering for State-of-the-art SEO Digital Strategies by Using NLP and ML},
  author={Gjorgjevska, Emilija and Mirceva, Georgina},
  booktitle={2021 3rd International Congress on Human-Computer Interaction, Optimization and Robotic Applications (HORA)},
  pages={1--6},
  year={2021},
  organization={IEEE}
}

@inproceedings{ohsaka2018boosting,
  title={Boosting PageRank scores by optimizing internal link structure},
  author={Ohsaka, Naoto and Sonobe, Tomohiro and Kakimura, Naonori and Fukunaga, Takuro and Fujita, Sumio and Kawarabayashi, Ken-ichi},
  booktitle={International Conference on Database and Expert Systems Applications},
  pages={424--439},
  year={2018},
  organization={Springer}
}

@misc{US9165040B1,
  author       = {Nissan Hajaj},
  title        = {Producing a ranking for pages using distances in a web-link graph},
  number       = {US9165040B1},
  year         = {2015},
  month        = oct,
  day          = {20},
  filed        = {2006-10-12},
  assignee     = {Google Inc.},
  note         = {Term extended by 268 days under 35 U.S.C. 154(b)},
}

@article{csaji2014pagerank,
  title={PageRank optimization by edge selection},
  author={Cs{\'a}ji, Bal{\'a}zs Csan{\'a}d and Jungers, Rapha{\"e}l M and Blondel, Vincent D},
  journal={Discrete Applied Mathematics},
  volume={169},
  pages={73--87},
  year={2014},
  publisher={Elsevier}
}

@misc{US8577893B1,
  author       = {Anna Patterson and Paul Haahr},
  title        = {Ranking based on reference contexts},
  number       = {US8577893B1},
  year         = {2013},
  month        = nov,
  day          = {5},
  filed        = {2004-03-15},
  assignee     = {Google Inc.},
  note         = {Priority and filing both on 2004-03-15; expected expiry March 28, 2032},
}

@misc{US8407231B2,
  author       = {Anurag Acharya and Matt Cutts and Jeffrey Dean and Paul Haahr and Monika Henzinger and Steve Lawrence and Karl Pfleger and Simon Tong},
  title        = {Document scoring based on link-based criteria},
  number       = {US8407231B2},
  year         = {2013},
  month        = mar,
  day          = {26},
  filed        = {2010-10-01},
  assignee     = {Google Inc.},
  note         = {Expired -- fee related; priority application filed Sept 30, 2003},
}

@misc{US20080134015A1,
  author       = {Natasa Milic-Frayling and Eduarda Mendes Rodrigues and Shashank Pandit},
  title        = {Website structure analysis},
  number       = {US20080134015A1},
  type         = {Patent application},
  year         = {2008},
  month        = jun,
  day          = {5},
  filed        = {2006-12-05},
  assignee     = {Microsoft Corporation},
  note         = {Application publication; priority filing Dec 5, 2006},
}

@inproceedings{baeza2005pagerank,
  title={Pagerank Increase under Different Collusion Topologies},
  author={Baeza-Yates, Ricardo A and Castillo, Carlos and L{\'o}pez, Vicente and Telef{\'o}nica, C{\'a}tedra},
  booktitle={AIRWeb},
  volume={5},
  pages={25--32},
  year={2005}
}

@inproceedings{fetterly2004spam,
  title={Spam, damn spam, and statistics: Using statistical analysis to locate spam web pages},
  author={Fetterly, Dennis and Manasse, Mark and Najork, Marc},
  booktitle={Proceedings of the 7th International Workshop on the Web and Databases: colocated with ACM SIGMOD/PODS 2004},
  pages={1--6},
  year={2004}
}

@article{avrachenkov2004decomposition,
  title={Decomposition of the google pagerank and optimal linking strategy},
  author={Avrachenkov, Konstantin and Litvak, Nelly},
  journal={INRIA Research Report},
  year={2004},
  publisher={INRIA}
}

@article{brin1998anatomy,
  title={The anatomy of a large-scale hypertextual web search engine},
  author={Brin, Sergey and Page, Lawrence},
  journal={Computer networks and ISDN systems},
  volume={30},
  number={1-7},
  pages={107--117},
  year={1998},
  publisher={Elsevier}
}

@misc{yandex_ranking_factors,
  title        = {Yandex Ranking Factors},
  url          = {https://yandex-ranking-factors.netlify.app/},
  urldate      = {2025-08-02}
}

@misc{wordliftCreatingInternal,
	author = {wordLift},
	title = {Creating Internal Links | WordLift Developer Documentation --- docs.wordlift.io},
	howpublished = {\url{https://docs.wordlift.io/agent-wordlift/workflows/create-internal-links/}},
	year = {},
	note = {[Accessed 03-09-2025]},
}

@article{barabasi1999emergence,
  title={Emergence of scaling in random networks},
  author={Barab{\'a}si, Albert-L{\'a}szl{\'o} and Albert, R{\'e}ka},
  journal={Science},
  volume={286},
  number={5439},
  pages={509--512},
  year={1999},
  publisher={American Association for the Advancement of Science}
}

@techreport{page1999pagerank,
  title={The PageRank citation ranking: Bringing order to the web.},
  author={Page, Lawrence and Brin, Sergey and Motwani, Rajeev and Winograd, Terry},
  year={1999},
  institution={Stanford infolab}
}

@inproceedings{lewandowski2021influence,
  title={The influence of search engine optimization on Google's results: A multi-dimensional approach for detecting SEO},
  author={Lewandowski, Dirk and S{\"u}nkler, Sebastian and Yagci, Nurce},
  booktitle={Proceedings of the 13th ACM Web Science Conference 2021},
  pages={12--20},
  year={2021}
}

@article{malaga2008worst,
  title={Worst practices in search engine optimization},
  author={Malaga, Ross A},
  journal={Communications of the ACM},
  volume={51},
  number={12},
  pages={147--150},
  year={2008},
  publisher={ACM New York, NY, USA}
}

@book{ledford2015search,
  title={Search engine optimization bible},
  author={Ledford, Jerri L},
  volume={584},
  year={2015},
  publisher={John Wiley \& Sons}
}

@online{mordor_seo_market_2025,
  title        = {SEO Market Size \& Share Analysis – Growth Trends \& Forecasts (2025–2030)},
  author       = {Mordor Intelligence},
  year         = {2025},
  url          = {https://www.mordorintelligence.com/industry-reports/seo-market},
  note         = {Accessed: 2025-09-29}
}

@article{khoshraftar2024survey,
  title={A survey on graph representation learning methods},
  author={Khoshraftar, Shima and An, Aijun},
  journal={ACM Transactions on Intelligent Systems and Technology},
  volume={15},
  number={1},
  pages={1--55},
  year={2024},
  publisher={ACM New York, NY}
}

@article{kipf2016semi,
  title={Semi-supervised classification with graph convolutional networks},
  author={Kipf, Thomas N and Welling, Max},
  journal={arXiv preprint arXiv:1609.02907},
  year={2016}
}

@article{velivckovic2017graph,
  title={Graph attention networks},
  author={Veli{\v{c}}kovi{\'c}, Petar and Cucurull, Guillem and Casanova, Arantxa and Romero, Adriana and Lio, Pietro and Bengio, Yoshua},
  journal={arXiv preprint arXiv:1710.10903},
  year={2017}
}

@article{hamilton2017inductive,
  title={Inductive representation learning on large graphs},
  author={Hamilton, Will and Ying, Zhitao and Leskovec, Jure},
  journal={Advances in neural information processing systems},
  volume={30},
  year={2017}
}

@inproceedings{gerlach2021multilingual,
  title={Multilingual entity linking system for wikipedia with a machine-in-the-loop approach},
  author={Gerlach, Martin and Miller, Marshall and Ho, Rita and Harlan, Kosta and Difallah, Djellel},
  booktitle={Proceedings of the 30th ACM International Conference on Information \& Knowledge Management},
  pages={3818--3827},
  year={2021}
}

@inproceedings{wu2011future,
  title={Future link prediction in the blogosphere for recommendation},
  author={Wu, Shanchan and Raschid, Louiqa and Rand, William},
  booktitle={Proceedings of the International AAAI Conference on Web and Social Media},
  volume={5},
  number={1},
  pages={642--645},
  year={2011}
}

@inproceedings{lyu2018real,
  title={Real-time event-based news suggestion for Wikipedia pages from news streams},
  author={Lyu, Lijun and Fetahu, Besnik},
  booktitle={Companion Proceedings of the The Web Conference 2018},
  pages={1793--1799},
  year={2018}
}

\section*{Appendix A. GenAI Usage Disclosure}
Large Language Models (LLMs) were used as assistants during the preparation of this paper. 
Specifically, LLMs supported (i) coding assistance for prototype implementation and experimentation, and (ii) grammar correction, stylistic polishing, and improving the organization and flow of the text. 
All conceptual contributions, experimental design, analyses, and conclusions were performed by the authors, who take full responsibility for this work.

\end{document}